\documentclass{article}[15pt]

\usepackage{amsmath,epsfig,color,calc,rotating,graphics,ulem}
\usepackage{float}

\usepackage{natbib}
\bibpunct{(}{)}{;}{a}{}{,}

\begin{document}

\title{
Range-Limited Heaps' Law for Functional DNA Words
 in the Human Genome 
\vspace{0.2in}
\author{
Wentian Li$^{1,2}$, Yannis Almirantis$^3$,  Astero Provata$^4$\\
{\small \sl 1. Department of Applied Mathematics and Statistics, Stony Brook University, 
Stony Brook, NY, USA\footnote{Current address.}}\\
{\small \sl  2. The Robert S. Boas Center for Genomics and Human Genetics}\\
{\small \sl  The Feinstein Institutes for Medical Research,  Northwell Health, Manhasset, NY, USA}\\
{\small \sl  3. Theoretical Biology and Computational Genomics Laboratory, Institute of Bioscience and Applications}\\
{\small \sl  National Center for Scientific Research ``Demokritos", 15341 Athens, Greece }\\
{\small \sl 4. Statistical Mechanics and Dynamical Systems Laboratory, Institute of Nanoscience and Nanotechnology }\\
{\small \sl National Center for Scientific Research, ``Demokritos", 15341 Athens, Greece}\\
}
\date{\today}
}
\maketitle  % End title section
\markboth{\sl Li et al. }{\sl Li et al. }

\newpage

\begin{center}
Abstract
\end{center}

Heaps' or Herdan's law is a linguistic law describing the relationship
between the vocabulary/dictionary size (type) and word counts (token) to be a
power-law function.  Its existence in genomes with certain definition of
DNA words is unclear partly because the dictionary size in genome
could be much smaller than that in a human language.  We 
define a DNA word as a coding region in a genome that codes for a protein
domain.  Using human chromosomes and chromosome arms as individual samples, 
we establish the existence of Heaps' law in the human genome within
limited range.  Our definition of words in a genomic or proteomic context is different 
from other definitions such as over-represented k-mers which are much shorter in length.
Although an approximate power-law distribution of protein domain sizes
due to gene duplication and the related Zipf's law is well known, their
translation to the Heaps' law in DNA words is not automatic.
Several other animal genomes are shown herein also to exhibit range-limited  Heaps' law 
with our definition of DNA words, though with various exponents. 
When tokens were randomly sampled and sample sizes reach to
the maximum level, a deviation from the Heaps' law
was observed, but a quadratic regression in log-log type-token plot fits the
data perfectly.  Investigation of  type-token plot and its  regression coefficients could 
provide an alternative  narrative of reusage and redundancy of protein domains 
as well as creation of new protein domains from a linguistic perspective.

\vspace{0.5in}

{\bf Abbreviations: 
AIC: Akaike information criterion; 
CRCPD: coding regions that code for protein domains;
GC content: percentage of guanine and cytosine bases.}

\large

\section{Introduction}

\indent

The study of molecular biology is full of examples in terminology, and in
quantitative and empirical laws, borrowed  from linguistics \citep{kay}.
Jargons used in molecular biology, such as transcription, translation, code, etc. are
often borrowed from terms in linguistics, not to mention the ubiquitous use of letters
(nucleotides and amino acids) and text (DNA and protein sequences). The recent Nobel-prize-winning
biotechnique of CRISPR was described as an equivalence of a text editor \citep{nelson}.
Not surprisingly, linguistic laws have also seen their mentioning in genomics.
For example, Zipf's law \citep{zipf35} which describes how the word frequency
in a text drops off with its rank, was tested in DNA sequences with k-mers/k-tuples 
(e.g. k=6) considered as ``word" \citep{mantegna}. This study was criticized, 
putting aside the issue of the
definition of DNA words, because the rank-frequency plot did not show a good power-law
as in the case of human languages \citep{konopka,li-zipf}. Another example of linguistic
law is the Menzerath's law \citep{menzerath}, which describes the tendency for 
using smaller linguistic units when these units are more numerous. The Menzerath's 
law at the human exon level is the tendency for genes with more exons to have 
smaller exon sizes \citep{li-menzerath,nikolaou}, while other choices of 
units are also investigated \citep{semple}. 

In this study, we focus on another linguistic law, the Heaps law \citep{heaps} 
(or Herdan-Heaps law \citep{egghe}), which states that the number of unique words 
(vocabulary/dictionary size) as a function of text length is a power-law. 
Heaps' law can be understood as a (number of)type-(number of)token 
\citep{tt-herdan,tt-wetzel} relationship: type being a unique word, token 
being usage of a word, i.e., the total number of types is the vocabulary 
size, and total number of tokens is the vocabulary usage.  Heaps' law can 
also be understood as a law of diminishing return \citep{petersen}. 
If one searches longer and longer texts for the purpose of
finding new words, Heaps' law indicates that the chance of finding new words
becomes smaller and smaller as more texts are looked over. The latter meaning
in Heaps law has been implied in the number of genes or gene-families discovered  
by sequencing more and more species' genomes in a clade \citep{medini-bc},
or the number of single-nucleotide-polymorphisms (SNPs) discovered by sequencing
more samples \citep{ionita}.

In order to apply a linguistic law to genomics, one first needs to find
an entity corresponding to a word. A common practice is to consider (overlapping) k-mers
with a prefixed k value as ``words" due to its easiness to calculate, its use in
alignment-free sequence comparison \citep{ziel,wli19}, and its importance in many
bioinformatics algorithms  \citep{chikhi,koren,rahman}.
However, the problem with all fixed-length  k-mers as words is  
that k-mers in general lack functional meaning. 

Instead of all k-mers, attempts were made to select
only the over-represented k-mers, based on the frequency expectation
from (k-1)-mers, as words or motifs \citep{brendel86,phillips87,vilo-thesis,apo03,gatherer}.
There are other motif detection methods based on statistical mechanics
\citep{bussemaker,mog}. However, these are postulated units only.
In fact, non-coding regions tend to have relatively more over-represented oligonucleotides
than coding regions \citep{frontali}. Many of over-represented
oligonucleotides in noncoding regions may be due to
error accumulation (e.g. slippage errors) tolerated by purifying 
selection, but some others may have functional implications and thus deserve the name 
``words", as the mirror-symmetrical words found in introns \citep{brendel86}. 

In the framework of semiotics \citep{peirce}, word is part of the signifier-signified pair
that represents a meaning. Without a biological meaning, a proposed DNA or protein word 
would be much less interesting if we continue to discuss linguistic laws.
Although the in-frame non-overlapping 3-mers within exons (codons) do have a 
biological meaning \citep{som}, k-mers,  even the
over-represented ones, may or may  not.

Recently, large language models \citep{llm}, such as bidirectional encoder representations
from transformers (BERT) \citep{bert} or  generative pre-trained transformer 
(GPT) \citep{radford}, have been applied to biomolecular sequences, in particular protein 
sequences \citep{rao,madani,nijkamp}. One aspect of natural language process is tokenization
which means to delineate a character string by boundaries \citep{webster}. For protein
sequences, the tokenization is a process to find ``words" \citep{ofer}. The protein ``words"
found by this method are usually short, with average size of only a few (e.g. four)
amino acids \citep{ferruz,dotan}. 

In this study, we use DNA segments that code a protein domain \citep{ywang} as our DNA 
``word tokens".  Treating protein domains as protein words has been proposed before 
\citep{searls,gimona,scaiewicz,yu,buchan}. 
There were studies that consider
gene internal structure in DNA sequences as ``grammar" or rules \citep{dong-searls},
but the emphasis of these were not on the definition of words. 
We prefer  to define DNA words as substrings of gene sequences
delimited by the boundaries of the protein domains within the proteins encoded
by those genes.
By doing so, our definition of DNA words has a biological meaning,
i.e. coding for a protein domain.

To investigate if Heaps' law holds in the human genome, our task is to find
all Coding Regions that Code for a Protein Domain (CRCPD).  For this we 
need to, for any genomic region, count the number of all CRCPDs (number of tokens)
and the number of distinct CRCPDs among them (number of types). 
For one chromosome of the  human genome, this 
pair of counts (\#token, \#type) will only give us one point. To see a relationship
between token and type, we repeat the same counting process for each one of
the 23 chromosomes and/or each one of the chromosome arms, so more points
are produced. This may be chosen if more points are desired, as
chromosomal arms may be seen as relatively independent entities,
and different from the whole chromosomes where they belong.

It has been known for over twenty years that the histogram of protein domains
or distinct parts follow an inverse power-law distribution in genomes
\citep{qian01,luscombe02,koonin02}. 
Another well known property is 
the importance role that gene duplication 
plays in shaping the distribution of these protein domains
as well as other genomic units \citep{muller,miller11,wli16}, echoing theoretical studies linking duplication and
power-law behavior \citep{wli91,ispolatov}. A perfect
inverse power-law histogram (distribution) will lead to a similar 
inverse power-law with rank as the x-variable, i.e., the Zipf's law \citep{newman}. 

However, more and more studies show that when tokens
are randomly chosen from a dictionary with a finite size whose words
follow Zipf's law distribution, the resulting type-token plot does
not follow an exact Heaps' law \citep{bernhardsson,font}.
In particular, the type-token plot in log-log scale would show
a concave downward shape \citep{font}. Nevertheless, within a limited range,
this concave curve might be approximately viewed as  a straight line.  
Also, publications with proofs on relations between two
power-law exponents in the two laws rely on certain assumptions,
e.g. infinite dictionary size, and working in the infinite token limit
\citep{boytsov}. All these observations imply that the power-law
distribution in protein domain known for the last twenty years
does not automatically lead to a power-law type-token plot for CRCPD,
which needs to be checked from the data directly.

In the next section, we describe the data and the methodology used.
In section 3, we present evidence of the existence of range-limited
Heaps' law in the human genome; while we point out one chromosome 
(chromosome 19) as an outlier to the Heaps' law; 
We also include in our study another level of genomic entities, 
chromosome arms, to explore the extent of Heaps' law. 
In section 4, we examine the potential Heaps' law in other animal genomes. 
In section 5, we generate artificial units by
randomly sampling DNA word tokens from the list of known DNA word
types, and confirm the previous results of log-log concave downward and
further reconfirm a systematic deviation from the Heaps' law.
The same random sampling process is also carried out at
the individual chromosome level, showing a distinction between
the concavity and finite dictionary size effect. 
Finally, various points are discussed in the Discussion
section, including a comparison between the
creation of new words in human languages and evolution  of
protein domains in genomes.

\section{Data and Methods}

The protein domains used in the present study refer to Pfam 
\citep{pfam20} (https://pfam.xfam.org/ or https://pfam-legacy.xfam.org/, 
which is hosted by InterPro \citep{interpro} after January 2023:
https://www.ebi.ac.uk/interpro).
The chromosomal locations of CRCPD in human genes are available
from the University of California at Santa Cruz (UCSC) Genome Browser track: 
Pfam domains in GENECODE genes. 
A description of
the methods used to derive this track can be found on UCSC Genome Browser site
({\sl https://genome.ucsc.edu/cgi-bin/hgTrackUi?db=hg19\&g=ucscGenePfam}).

To obtain this Pfam domains track in the human genome (for GRCh38/hg38),
from the top bar of the genome browser, we select Tools $\rightarrow$ Table Browser, 
group = ``Genes and Gene Predictions",
track = ``Pfam in GENECODE". The resulting table contains chromosome number,
chromosomal location (start, end, in bp), name of the Pfam domain, etc.

In this study, we also used chromosome arms as units.
To determine the chromosome arm to which a region belongs to, we download
the cytoband information from: \\
https://hgdownload.cse.ucsc.edu/goldenpath/hg38/database/cytoBand.txt.gz
to find the centromere locations in each chromosome.
The counting of Pfam domains (both token and type) was carried out by
R (https://www.r-project.org) scripts written by us.

The Heaps' law is written as ($y$= vocabulary size = \#type, or the number of unique  
CRCPD domains in a chromosome,  $x$=vocabulary usage=\#token, or number of times any of these 
CRCPD appear in the same chromosome):
\begin{equation}
\label{eq-heaps}
y = C x^\alpha
\end{equation}
which after a logarithmic transformation on both side ($y'=log(y)$, $x'=log(x)$, $C'=log(C)$):
$ y'= C' + \alpha x' $
becomes a linear relationship. The exponent $\alpha$ is determined by a linear
regression in R (the function {\sl lm}). 
For model selection the Akaike information criterion (AIC) 
\citep{aic,wli-aic} calculation is carried out by the AICcmodavg package 
({\sl cran.r-project.org/web/packages/AICcmodavg/}), using the function
{\sl aictab}. For quadratic regression, we fit the equation
$y'= C + \alpha x' + \beta x'^2$.

The following notation may gives readers an idea of the type of data 
we have. Suppose there 
are N tokens (appearance of CRCPDs: coding for Pfam domains), and K unique ones.
$N=\sum_{i=1}^K P_k$, where $P_k$ is the multiplicity of word type-$k$.
For any genomic region (chromosome, chromosome arm, or half-arm), we count
the number of tokens $x_i$ ($i$ is the index for sample point) and $y_i$
of them are unique. When we have enough samplings $i=1,2, \cdots, I$,
these ($x_i$, $y_i$) pairs can be log-transformed and  used for a linear regression.
We may not be constrained by genomic regions, but sample $x_i$ tokens genome-wide
from the pool of N tokens (sampling without replacement) and among them $y_i$ ones
are unique. This can be repeated I times with ever larger $x_i$ values.
The same ($x_i$,$y_i$) pairs can be used to check Heaps' law.

\section{Results for the human genome}

\subsection{Heaps' law for Pfam CRCPD holds true}

The number of Pfam domain appearance, 
as our example of CRCPD, in each autosomal and X chromosome
in the human genome (version hg38) is obtained from the Pfam-in-GENECODE track
of the UCSC genome browser. This would produce 23 points for checking the
vocabulary size (type) and vocabulary usage (token) relationship and a regression.  
In order to increase the sample size, 
we also consider data from the p- and q-arm in a chromosome, as partitioned by the centromere
for each metacentric (non-acrocentric) chromosome (41 points).

\begin{figure}[H]
 \begin{center}
 \includegraphics[width=0.8\textwidth]{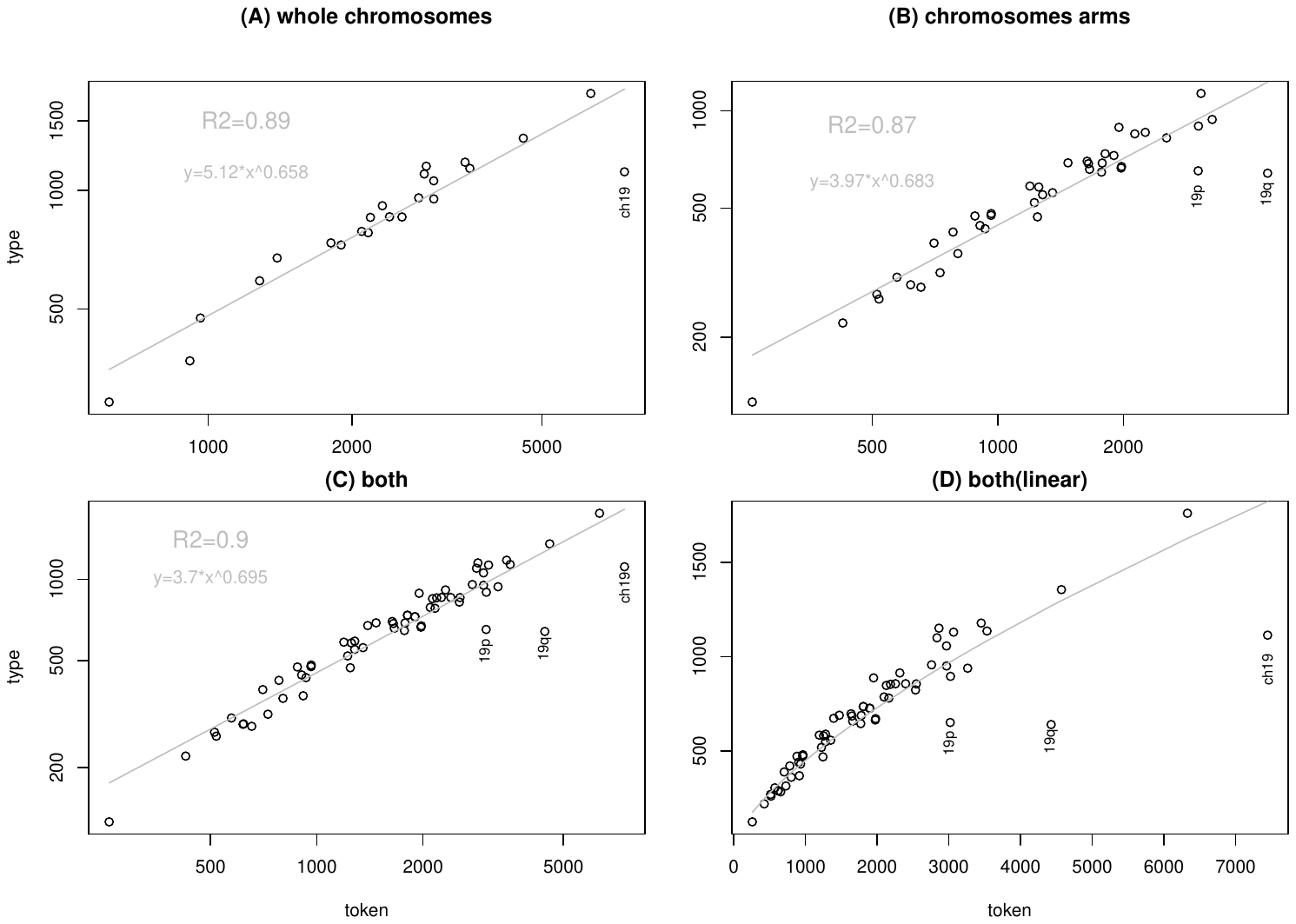}
 \end{center}
\caption{ \label{fig1}
Type-token plot, where $x$-axis is the number of CRCPD tokens
(vocabulary usage) and $y$-axis is the number of unique CRCPD type
(vocabulary size), for human genome. (A) Each point corresponds to a human chromosome,
in log-log scale.  (B) Each point corresponds to a human chromosome arm, in log-log scale. 
(C) Each point corresponds to either a human chromosome or a chromosome arm, in log-log scale. 
(D) Same as  (C) but in linear-linear scale.
Chromosome 19 is clearly indicated as an outlier in all panels.
}
\end{figure}

Fig.\ref{fig1}(A) shows the number of types (number of unique/distinct Pfam domains)
as a function of number of tokens (Pfam domain counts) for 23 human chromosomes
in log-log scale. With the exception of one point, all other points scatter 
around a straight-line, indicating a power-law and Heaps' law. The outlier
is chromosome 19, being lower in number of CRCPD types or higher in
number of CRCPD  tokens.  The regression for log-token over log-type produces the
slope (scaling exponent) $\alpha=0.658$ and $C=5.12$ in Eq.(\ref{eq-heaps}). 

Fig.\ref{fig1}(B) shows a similar type-token plot in log-log scale using
chromosome arm data. The regression results are similar: $\alpha=0.683$ and $C=3.97$.
Using both whole chromosome and chromosome arm data leads to Fig.\ref{fig1}(C),
with $\alpha=0.695$ and $C=3.7$. Figure \ref{fig1}(D) shows the same data
as Fig.\ref{fig1}(C) but in linear-linear scale. It is clear that  power-law
curve function fits the data better than a linear regression. In fact, 
the variance explained by regression is much higher in log-log scale 
($R^2=0.887, 0.873, 0.9$ for chromosome data, chromosome arm data, and 
combined) than in linear-linear scale ($R^2=0.697, 0.667, 0.738$).

Removing the outliers (one point in Fig.\ref{fig1}(A), two points in Fig.\ref{fig1}(B)
for two chromosome 19 arms, and three points in Fig. \ref{fig1}(C)) result
in even better $R^2$: 0.968, 0.955, and 0.967. The slope of regression
without outliers is up slight: $\alpha=0.754, 0.772, 0.770$, and $C$ is down to
around 2.

\subsection{The protein domains that are  mostly responsible for the redundancy} 
Apparently, more tokens than types means
that some CRCPDs appear multiple times in a chromosome. Which
CRCPDs  appear more than others? Table \ref{table1}
shows the top three Pfam domains in each human chromosome, as well as in the  whole genome,
according to their frequency. On the whole genome level, zinc finger C2H2 type (zf-C2H2),
a C2H2-type zinc finger (zf-C2H2\_6), and Immunoglobulin I-set domain (I-set) are
the most frequent domains, appearing 5493, 1609, and 900 times, respectively.
Genomewide there are 6839 Pfam domain names, whose ID and annotation can be
found in
http://ftp.ebi.ac.uk/pub/databases/Pfam/current\_release/Pfam-A.hmm.gz
or http://pfam-legacy.xfam.org/family/browse .

On individual chromosome level, zf-C2H2 domain \citep{miller} is among the top 3 domains
for 19 out of 23 chromosomes (not on the top 3 for  chromosomes 2,11,14, and 22).
Other top ranking domains on individual chromosome level include
olduvai domain (Chr1), immunoglobulin I-set domain (Chr2), cadherin domain (Chr4,5),
7 transmembrane receptor (rhodopsin family) (Chr11), collagen triple helix repeat 
(Chr13), immunoglobulin V-set domain (Chr14, 22), and keratin high sulfur B2 protein (Chr21).

\begin{table}
\begin{center}
\begin{tabular}{c|ccccccc}
\hline
ch & most\_freq CN & 2nd CN & 3rd CN & num\_type & per(CN=1) & mean(CN) & max(CN) \\
\hline
1  & Olduvai 215  & zf-C2H2 214  & Sushi 186  &  {\bf 1759}  &  64.7\%  &  3.597  &  215 \\
2  & I-set 224  & fn3 156  & Ig\_3 136  &  1355  &  63.8\%  &  3.373  &  224 \\
3  & zf-C2H2 217  & WD40 72  & Ig\_3 70  &     1178  &  65.3\%  &  2.93  &  217 \\
4  & Cadherin 101  & zf-C2H2 92  & Ank 50  &   857  &  67.1\%  &  2.797  &  101 \\
5  & Cadherin 334  & zf-C2H2 134  & Cadherin\_2 55  &  951  &  68.2\%  &  3.121  &  334 \\
6  & zf-C2H2 162  & fn3 62  & zf-C2H2\_6 57  &  1057  &  65.8\%  &  2.807  &  162 \\
7  & zf-C2H2 324  & zf-C2H2\_6 91  & V-set 58  &  957  &  66.1\%  &  2.883  &  324 \\
8  & zf-C2H2 171  & Sushi 56  & Ank 52  &  787  &  67.9\%  &  2.663  &  171 \\
9  & zf-C2H2 144  & I-set 57  & Ig\_3 57  &  856  &  67.4\%  &  2.974  &  144 \\
10  & zf-C2H2 103  & Ank 72  & Ank\_3 44  &  854  &  66\%  &  2.56  &  103 \\
11  & 7tm\_1 196  & 7tm\_4 180  & I-set 70  &  1136  &  65.8\%  &  3.109  &  196\\
12  & zf-C2H2 127  & WD40 48  & Ank 40  &  1100  &  65.8\%  &  2.576  &  127 \\
13  & Collagen 28  & Cadherin 21  & zf-C2H2 19  &  475  &  69.7\%  &  {\it 2.027}  &  {\it 28} \\
14  & V-set 96  & WD40 41  & 7tm\_1 40  &  736  &  67.1\%  &  2.455  &  96 \\
15  & zf-C2H2 57  & EGF\_CA 46  & hEGF 41  &  727  &  68.1\%  &  2.611  &  57 \\
16  & zf-C2H2 292  & zf-C2H2\_6 71  & WD40 44  &  914  &  67.2\%  &  2.534  &  292 \\
17  & zf-C2H2 107  & Keratin\_B2\_2 73  & WD40 42  &  1151  &  65.2\%  &  2.487  &  107 \\
18  & zf-C2H2 75  & Cadherin 45  & Laminin\_EGF 30  &  370  &  69.5\%  &  2.473  &  75 \\
19  & zf-C2H2 2805  & zf-C2H2\_6 910  & KRAB 228  &  1114  &  65.4\%  & {\bf 6.684}  & {\bf 2805} \\
20  & zf-C2H2 107  & zf-C2H2\_6 25  & Laminin\_EGF 21  &  674  &  {\bf 72.6\%}  &  2.068  &  107 \\
21  & Keratin\_B2\_2 47  & I-set 17  & ig 16  & {\it 291}  &  69.4\%  &  2.131  &  47 \\
22  & V-set 39  & zf-C2H2 27  & EGF\_CA 16  &  590  &  68.1\%  &  2.171  &  39 \\
X & zf-C2H2 136  & Collagen 32  & MAGE 32  &  781  &  {\it 59.3\%}  &  2.77  &  136\\
\hline
all & zf-C2H2 5493 & zf-C2H2\_6 1609 &     I-set 900 & 6839 & 44.3\% & 9.109 & 5483 \\
\hline
\end{tabular}
\end{center}
\caption{
\label{table1}
Some human chromosome level statistics on protein domains (CN: copy number,
or number of CRCPD token for a specific CRCPD type):
the protein domain with the highest, second highest, and the third highest, CN;
number of unique CRCPDs;
proportion of CRCPD types that have CN=1 (which could be called a singleton);
average copy number (or token-over-type ratio);
and maximum copy number. The maximum (minimum) value in each column is marked with bold (italics). 
}
\end{table}

\subsection{Chromosome 19 as outlier}

Chromosome 19 (Chr19) has already been known to have the highest gene density
compared to other chromosomes, more gene families, high GC content, higher
density of repetitive sequences \citep{grimwood}, extreme divergence with mouse
genome \citep{castresana} but conserved within nonhuman primates \citep{harris}. 

We examine several chromosome level statistics to see which quantities make
chromosome 19 an outlier. Chr19 has the highest GC content compared to
other chromosomes, which might be related to the higher gene density,
which itself is a product of widespread DNA/gene duplication.

If CRCPD statistics are examined directly, we find that the
proportion of unique CRCPD  
in Chr19 does not stand out (see Table \ref{table1}). The
number of CRCPD types on Chr19 is large, but not the largest. 
However, Chr19 has the highest average copy number per CRCPD, and also
has the largest copy number for a CRCPD:
zinc finger C2H2 domain appears 2805 times, which is an order of magnitude
larger than the maximum copy number in the other chromosomes.

\subsection{Extension to half of the chromosome arms} 
\begin{figure}[H]
 \begin{center}
  \includegraphics[width=0.8\textwidth]{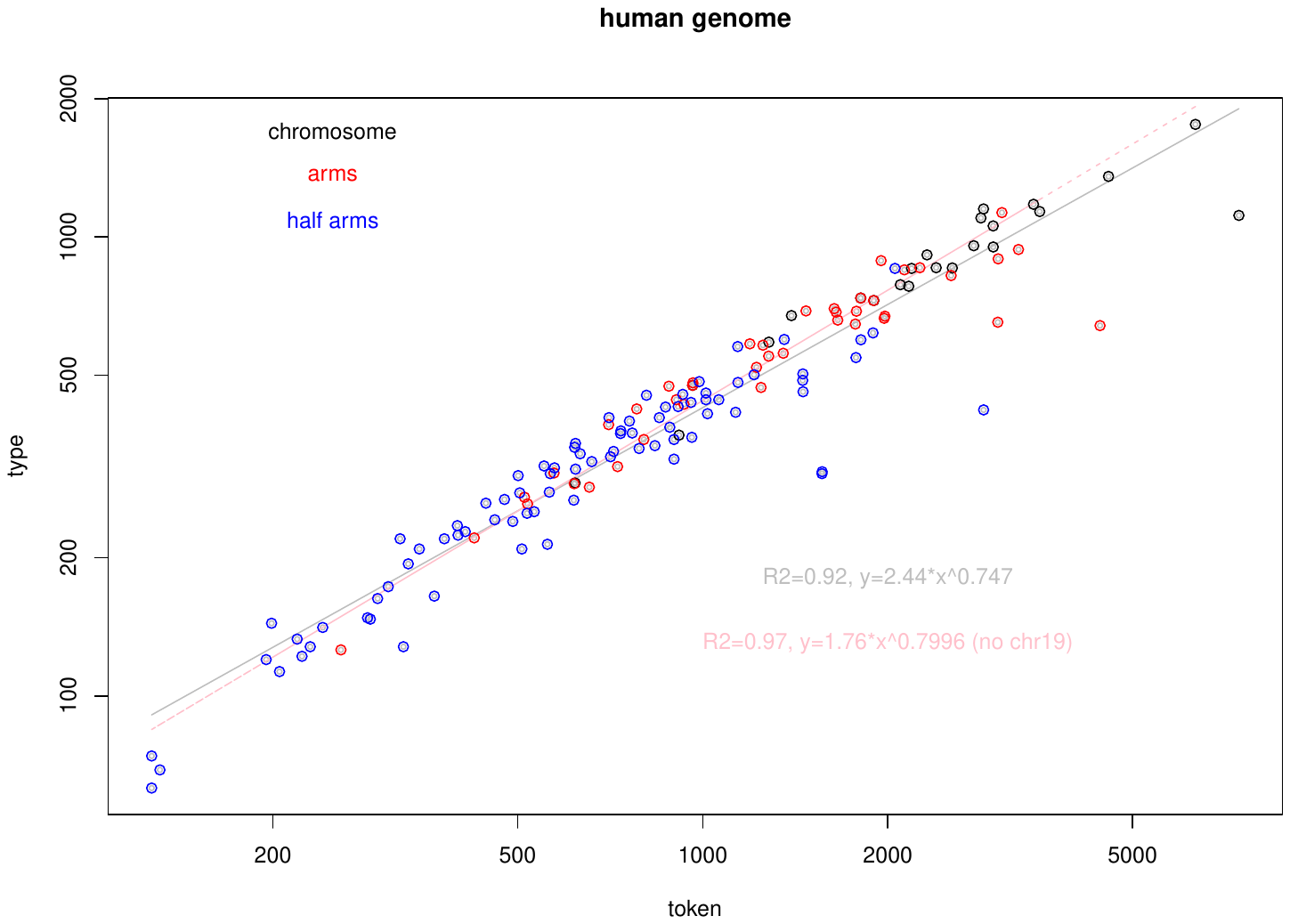}
 \end{center}
\caption{ \label{fig2}
An extension of Fig.\ref{fig1}(C). Besides human chromosomes and chromosome
arms, half-chromosome-arms (blue) are used to count CRCPD tokens and types.
}
\end{figure}

The x-axis of Fig.\ref{fig1}
only spans one decade (ratio of the maximum and minimum token count
is 12 for chromosomes and 17 for chromosome arms). In order to increase
the range of x-axis, we split each chromosome arm into two equally sized
regions. The type-token with this relatively local data being added is shown in
Fig.\ref{fig2}.

Figure \ref{fig2} shows that with more data at the low end of the x-axis
(the largest token count is 59 times the smallest token count), Heaps'
law remains to be true, with the fitting exponent $\alpha = 0.75$. 
Removing the outliers due to Chr19, $\alpha = 0.8$. Both of these
slopes are larger than those in Fig.\ref{fig1}. 

Visual inspection of Fig.\ref{fig2} seems to show a slight tendency 
for the points to curve down,
which might explain the increase of the fitting exponent value.
In fact, in the original version of Heaps' law in Eq.(\ref{eq-heaps}),
$C$ should be 1 as the line should go through $x=y=1$ points
(when there is only one token, it must be unique).  All our fitting $C$
values are larger than 1, meaning that in the log-log scale, type-token
scatter plot will eventually deviate from a straight line towards the origin.
We will return to this issue in section 5.

\section{Robustness of the results}

\subsection{Random shuffle of CRCPDs}
 
\indent

In order to check if the sampling units of chromosomes, chromosome arms,
or half arms, play any role in the observed power-law trend in Figs.\ref{fig1},\ref{fig2},
we carried out two experiments. The first is that we shuffle the CRCPDs (tokens)
genome-wide once, while keeping the chromosome location intact. Then we repeat the
same CRCPD-type count in the same chromosomes, chromosome arms, and half arms.  
In the next subsection, we will randomly sample CRCPDs without replacement.
Then the CRCPD-type count is plotted against the number of CRCPD token sampled. 

\begin{figure}[H]
 \begin{center}
  \includegraphics[width=0.8\textwidth]{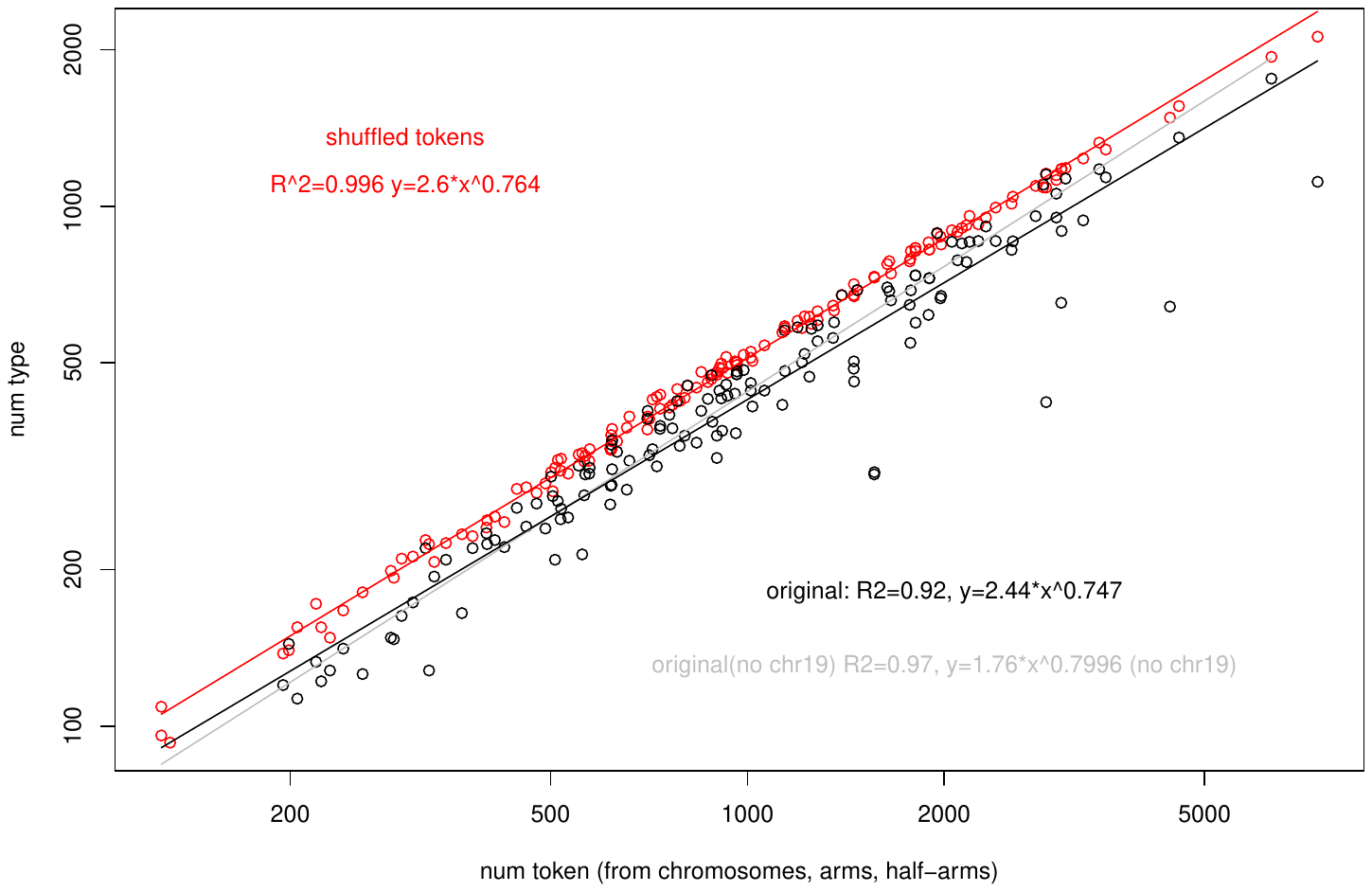}
 \end{center}
\caption{ \label{fig3-sec41}
CRCPD types count in chromosomes, chromosome arms,
and half arms after CRCPD tokens are shuffled genome-wide (red dots),
as a function of CRCPD token counts in these genomic units. 
The black dots are idential to those in Fig.\ref{fig2}, for
the same type-token counts when CRCPDs are not shuffled. 
}
\end{figure}

Fig.\ref{fig3-sec41} shows the type-token relationship when CRCPDs are
shuffled in the whole genome. The black dots are identical to those in
Fig.\ref{fig2}, some dots represent chromosomes, others represent chromosome
arms or half arms. The red dots are the similar type counts in the same
genomic segments (chromosomes, arms, half arms) when the shuffling has 
taken place.

It can be seen from Fig.\ref{fig3-sec41} that after shuffling, each genomic
unit tends to have more CRCPD types than the original data (red dots being
above black dots). It means that CRCPD redundancy tends to be locally
clustered. However, the shuffling of CRCPDs does not change the fact
that type and token are in a power-law relationship.
The regression coefficient is 0.747 for the shuffled data, very close 
to the original values of 0.747 (or 0.8 when Chr19 is removed)
for the original data.

\subsection{Random sampling from CRCPD pool 
and a systematic deviation from the power-law}
 
\indent

\begin{figure}[H]
 \begin{center}
 \hspace*{-1in}	
  \includegraphics[width=0.7\textwidth]{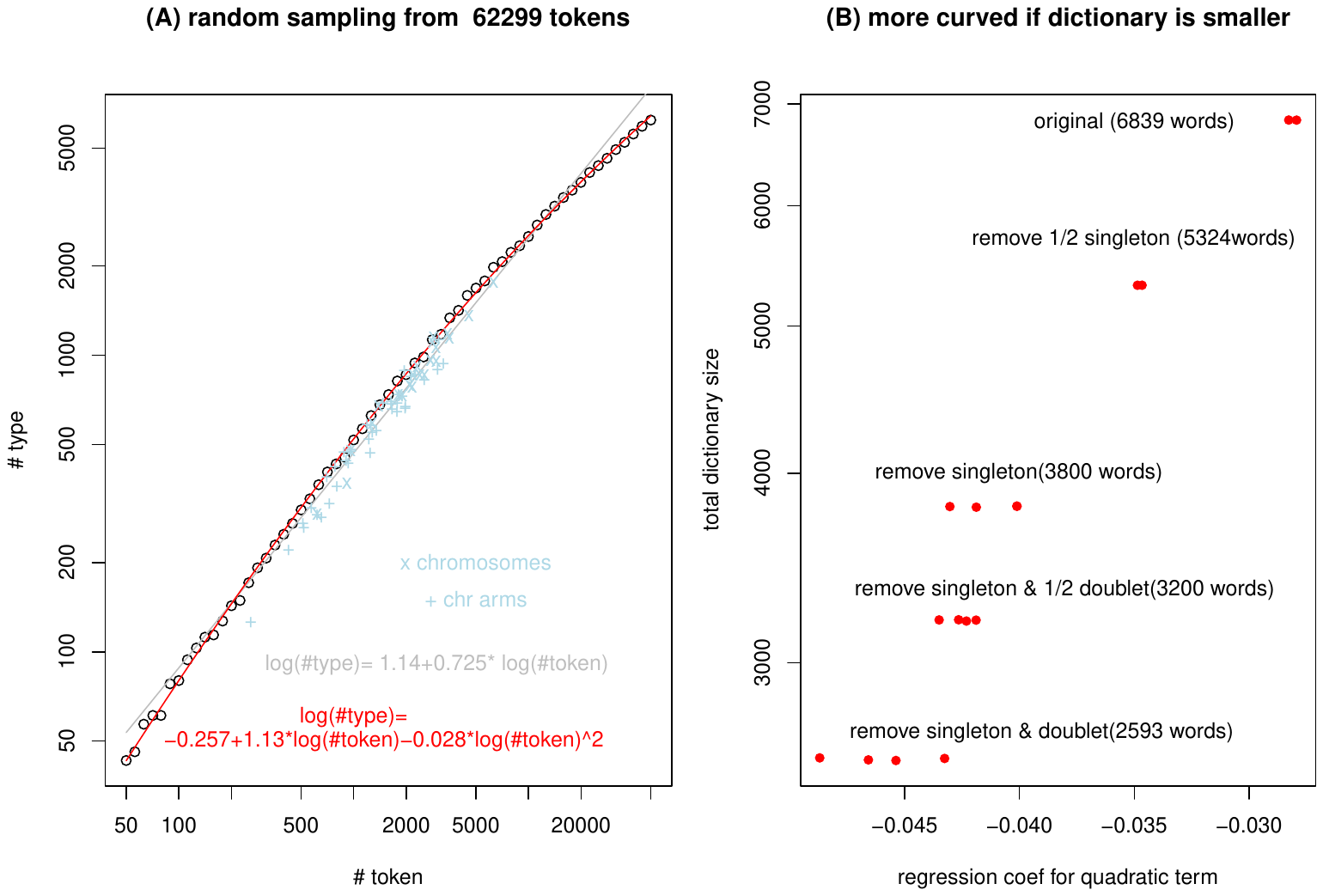}
 \end{center}
\caption{ \label{fig4-random}
(A) Token(x-axis)-type(y-axis) plot for randomly generated token
from the whole token pool of 62299 CRCPDs. The plot is fitted
by both linear regression (after logarithm transformation of both axes) and
quadratic regression. Quadratic regression visibly fits the data better than
linear regression, indicating a deviation from the Heaps' law.
The data points from Fig.1(C) are redrawn here (lightblue) for comparison
(for both chromosomes and chromosome arms, except for Chr19). 
(B) Repeating the same random sampling of token and quadratic regression
fitting, when the dictionary size is gradually reduced. 
The y-axis shows the number of CRCPD type (from top to bottom): original data;
half of the singleton CRCPD types (those that appear in the human
genome only once) are removed; all singletons are removed; all singletons and
half of the doublet CRCPD types are removed; all singletons and all
doublets are removed. The x-axis is the regression coefficient for the
quadratic term.
}
\end{figure}

In the second approach to test the robustness of our
result, we create artificial sampling units by randomly picking  CRCPD
tokens from the genome. This new way of sampling tokens would free
us from the restriction of chromosomes or any genomic regions. Also,
the range of x-axis in type-token plot can be greatly expanded. 
We pool CRCPDs from all human  chromosomes
(62299 tokens). Then we randomly sample x number of tokens and check how
many of them are unique (y number of types). A type-token plot from this random
sampling is shown in Fig.\ref{fig4-random}(A) in log-log scale.
Note that the x-range in Fig.\ref{fig4-random}(A) is much
more expanded compared to Figs.\ref{fig1},\ref{fig2}.

If Heaps' law holds exactly true, we should expect
a straight line in the log-log plot in Fig.\ref{fig4-random}. The linear regression
(grey line) captures this trend (exponent 0.725 is less than one).
However, a systematic deviation from the straight line, of a concave shape 
at both ends of the line, can be easily seen. When we add a quadratic term, 
it fits the data perfectly (with the coefficient for the quadratic term to 
be negative). 
The AIC of the quadratic regression, $-244.79$, is much better
than that for the linear regression ($-93.08$), confirming the visual impression.
This ``log-log concavity" was previously reported
\citep{bernhardsson,lv,font} and our DNA words result fully conforms it.
For this reason, a better term for plots in 
Figs.\ref{fig1},\ref{fig2},\ref{fig6-other},\ref{fig4-random}(A)
could be ``range-limited Heaps' law".

Since quadratic curves fit the type-token plot in Fig.\ref{fig4-random}(A) nicely,
the coefficient of the quadratic term is a reliable measure of the concavity.
We investigate the impact of limited dictionary size on the concavity,
or deviation from the Heaps' law,  by removing rare CRCPD.
Fig.\ref{fig4-random}(B) shows the fitting
coefficient of the quadratic term at these situations: half of the singleton
CRCPD types (those that appear in the genome only once) are removed;
all singleton CRCPD types are removed; all singleton plus half of
the doublet (those CRCPD types that appear in the genome only twice)
are removed; and all singletons and all doublets are removed.

Each point in Fig.\ref{fig4-random}(B) represents a random sampling
process similar to that in Fig.\ref{fig4-random}(A), i.e., a sequence
of token sampling followed by type counting, when the type source (dictionary)
is cut short. We ran the random sampling multiple times even with the
same dictionary size, because of the variation of random sampling, as well
as changing the upper limit of the x-axis to accommodate the reduction
of dictionary size. Even with these variations, the trend is clear that
the type-token plot becomes more curved if more rare words (singleton
and doublet protein domains) are removed.

\begin{figure}[H]
 \begin{center}
  \includegraphics[width=0.8\textwidth]{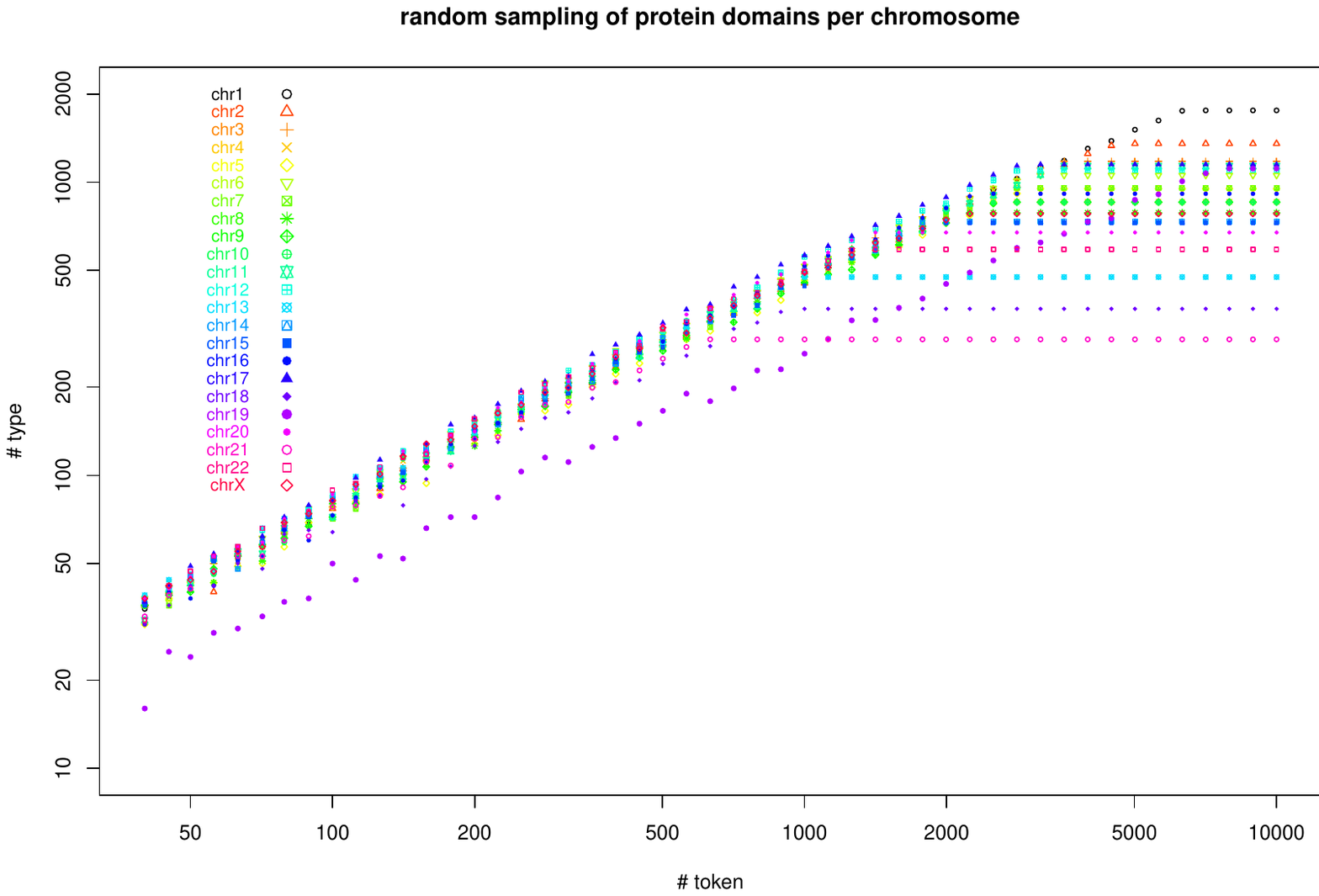}
 \end{center}
\caption{ \label{fig5}
Type-token plots for randomly sampled CRCPDs in individual chromosomes.
Note that Chr19 (lower purple curve) distinguishes itself from
the rest (see discussion in the text).
}
\end{figure}

The random sampling of tokens can also be carried out on per chromosome
basis. Fig.\ref{fig5} shows the type-token plot for 23 individual
chromosomes.  Protein domains observed only within one chromosome are
pooled, then randomly sampled without replacement, and distinct types
are counted. Chromosome 19 again stands out as the chromosome with
the lowest diversity and lesser number of types. The plateaus reached
for each chromosome indicate that the dictionary size has been reached.
This finite size effect is different from the slight convexity of the
type-token curve in Fig.\ref{fig4-random}. 

The lines/curves for individual chromosomes in Fig.\ref{fig5}
behave similarly in terms of slope before reaching the plateau.
The y-intercept is very different for chromosome 19, consistent
with Fig.\ref{fig1} and Fig.\ref{fig2}, while those of
other chromosomes are comparable. The shape of per-chromosome
type-token scatter plots obtained by random sampling, which is
roughly a power-law with a slight concave curving followed by a plateau, 
is a typical example of finite size effect. Due to the nature of generated
data, the number of tokens is not limited whereas the number of types is
finite, leading to plateaus.

\section{Results for other animal genomes}

\subsection{Sampling from chromosomal units}

\indent

Using the same UCSC Genome
Browser, we downloaded the Pfam domain locations for mouse genome (GRCm39/mm39, Jun 2020,
all 21 chromosomes), UniProt \citep{uniprot} domain information for chicken 
(GRCg6a/galGal6, 2018, missing
chromosomes 29, 34-39), zebrafish (GRCz11/danRer11, 2017, all 25 chromosomes), and
drosophila melanogaster (BDGP release 6, 2014, 2L, 2R, 3L, 3R, 4, X and Y chromosomes/arms).

\begin{figure}[t]
 \begin{center}
  \includegraphics[width=0.8\textwidth]{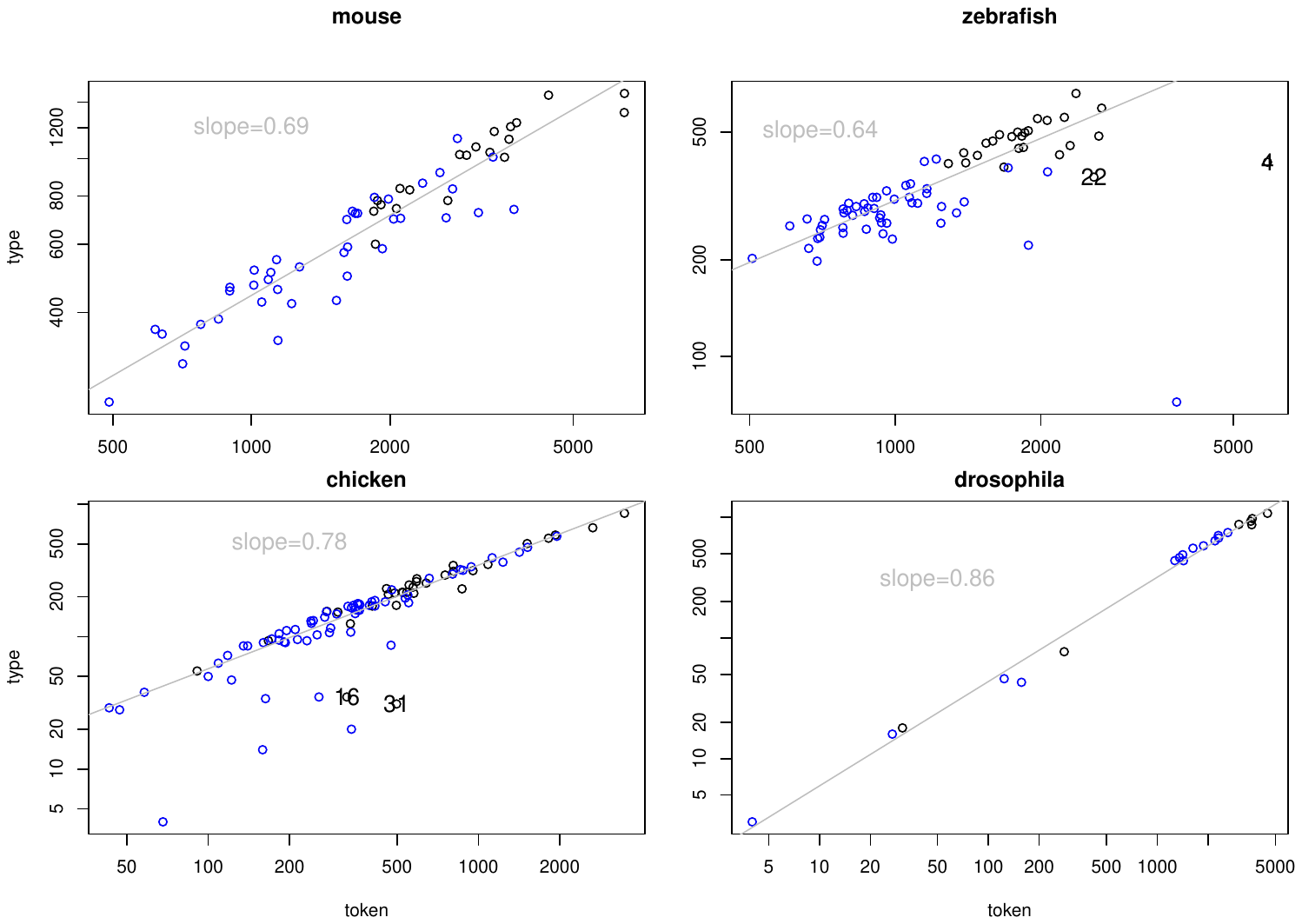}
 \end{center}
\caption{ \label{fig6-other}
Token-type plots for mouse, zebrafish, chicken, and drosophila genomes.
Each black dot represents a chromosome and blue dot for half-chromosome.
The chromosome number of some outliers are indicated. The regression
slopes from log-log plot are the Heaps' law exponent.
}
\end{figure}

Figure \ref{fig6-other} shows the type-token scatter plot in log-log scale
with points both representing a chromosome (black) or half of a chromosome (blue).
The practice of using half chromosomes is to increase the sample size
and to increase the covering range of the x-axis.

The mouse genome exhibits similar behavior as the human genome. The variance explained
by the power-law is $R^2=0.887$, the regression slope is $\alpha=0.688$ and $C=3.8$.
The zebrafish genome has two chromosome outliers, chromosome 4 and 22, 
as well as half chromosome from chromosome 4.  Removing these four points 
lead to a power-law regression with $R^2=0.816$ and slope=0.643, $C=3.62$.
The chicken genome has two outliers at the chromosome level: Chr16 and Chr31,
and 4 half-chromosome level outliers derived from these two chromosomes.
After removing these six points, the regression in log-log scale has
$R^2=0.937$, slope=0.782, and $C=1.56$. Finally, for drosophila genome,
$R^2=0.993$, slope=0.865, and $C=0.81$.

To summarize, the results from these genomes where the protein domain track
is available, Heaps' law is generally true with exponent $\alpha < 1$,
despite potentially data quality issues such as the UniProt domain information,
limited range of x-axis, and limited number of sample points. We also
notice that when the low-end of the x-axis for the scatter plot is close 
to the origin (e.g. drosophila), the slope is  relatively larger and closer to 1. 
Another observation is the difference of vocabulary sizes
among different genomes, with mouse having more protein domain types
than chicken, zebrafish and drosophila. Word size differences between genomes
were also observed in \citep{gatherer}, though for different definition
of words (over-represented protein substrings).

\subsection{Random sampling from total CRCPD pool}

\indent

Similar to subsection 4.2, we may 
create our own sampling units by randomly sampling CRCPD tokens from a genome. 
This step is even more important for other animal genomes than the human 
genome, because the type-token plots in Fig.\ref{fig6-other} are more 
noisy and are more limited in x-axis range. 

\begin{figure}[t]
 \begin{center}
  \includegraphics[width=0.8\textwidth]{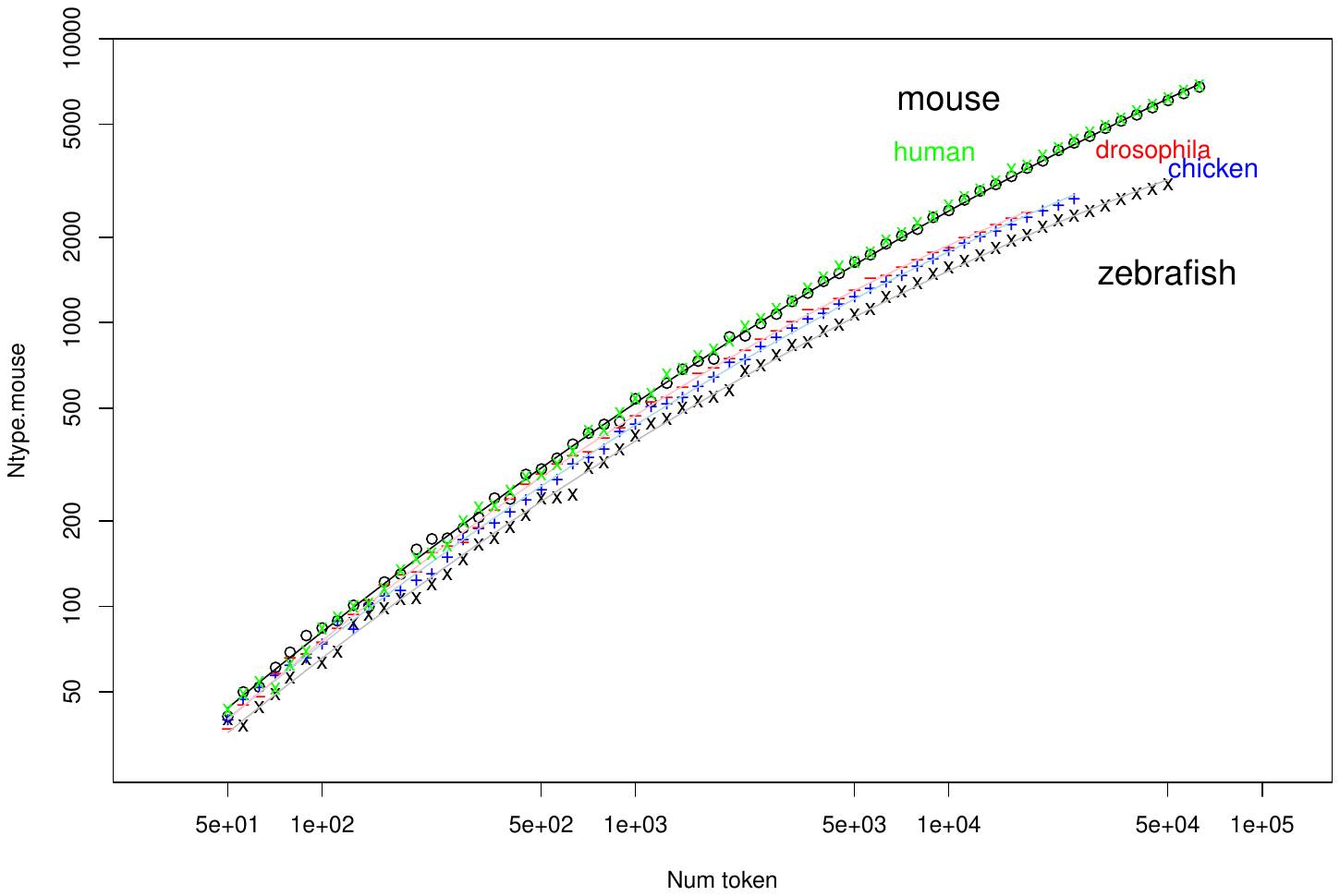}
 \end{center}
\caption{ \label{fig7-other-random}
For the four animal genomes (mouse, drosophila, chicken, and
zebrafish), number of unique CRCPD types as a function of the number
of randomly sampled CRCPD tokens (in log-log scale). The quadratic
regression lines (for log-x and log-y) are also shown.}
\end{figure}

Figure \ref{fig7-other-random} shows the number of
CRCPD types within a randomly sampled CRCPD tokens (without replacement),
as a function of number of tokens, for the four genomes (mouse, 
drosophila, chicken, and zebrafish). The trend of these type-token
plots is very similar: quadratic regression lines fit the data perfectly,
and within limited range, these quadratic lines can be approximated
as linear segments (thus Heaps' law). 

The quadratic nature of the type-token plot when
x-axis is extended reveals a weakness of using linear regression to
fit the data in Fig.\ref{fig6-other}. The slight difference between
the linear coefficients in the different genomes in Fig.\ref{fig6-other}
is not a reliable index to rank genomes. Figure \ref{fig7-other-random}, on
the other hand, shows a clear order, from zebrafish, to chicken and drosophila,
then to mouse, following the trend of increase in 
vocabulary sizes (number of distinct CRCPD types).
The human data is also shown in Fig.\ref{fig7-other-random} which is
essentially identical to that of mouse.

\section{Discussion}

\indent

Heaps (or Herdan-Heaps) law is one of the classic
linguistic ``laws" \citep{altmann,hernandez}. The word ``law" in this context is very different from
that in other fields such as physics (e.g. Newton's laws). Linguistic laws
are empirical, not supposed to be exact, and only describe a trend. 
As reported by several authors and confirmed in our own DNA word data,
Heaps' law or power-law relationship (or linear relation in log-log scale)
is unlikely to be true for all range of text lengths (number of tokens)
\citep{bernhardsson,lv,font}. Even for human languages
(e.g., some non-Indo-European languages), \#type tends to flatten out
away from the power-law in the large \# token end \citep{lv}. 
At low \#token end, the fact that the fitting parameter C $>$ 1 (or C' $>$ 0), in our data, 
indicates that the power-law will not be extrapolated all the way to the origin.

Since the genome size is limited, the total number of protein-domain-coding
loci is limited, therefore there is an upper limit in the \#token axis
as well as in the \#type axis. Similarly,
due to the large proportion ($\sim$ 50\%) of human genome
being in non-coding intergenic regions without any CRCPDs,
it is not practical to use a short genome region as it may have zero count
of \#token and \#type.
This leads to a lower limit in \#token axis.
Checking type-token relation in the human genome has to be range limited.
On the other hand, our simulation in Fig.\ref{fig4-random} shows that
beyond certain range, one should not expect power-law relationship between
type and token anymore, and Heaps' law will break down.

There are several publications relating exponents in
Zipf's law and Heaps' law  asymptotically {\sl if} both
the token frequency vs. rank and type vs. token are indeed inverse
power-law and power-law \citep{baeza}. There are also papers
deriving the asymptotic type-token relation in the model
of random sampling from a power-law distributed word pool
\citep{eliazar,gerlach,boytsov}, in complicated mathematical expressions 
often in the form of infinite summation. With finite dictionary
size, infinite summation becomes finite summation, but still
no simple mathematical expression is proposed. Here, from a data
fitting perspective, the type-token relation can be expressed
by a quadratic equation: $\log(y)=c' +\alpha \log(x) - \beta (\log(x))^2$
with $0 < \alpha < 1$ and $\beta > 0$. Using an extra term
and extra parameter \citep{wli-entropy,wli-letter},
in particular, using a quadratic/parabolic term \citep{frappat},
to correct deviation from a straight line is a common practice.

Studies along the same lines with the present investigation include
\citep{nasir17, caetano21}. These works focus on phylogenetics of
different species, and the word used is protein fold superfamilies (FSF).
FSF information was provided by the SCOP (structural classification of proteins)
database \citep{scop,scop20}.
Even if SCOP domain superfamilies and Pfam domains are somewhat similar,
the sample points in a type-token plot in \citep{nasir17, caetano21}
are different from those in the current study. In \citep{nasir17, caetano21},
each point represents a species whereas in our study a point is a chromosome or
chromosome arm or a unit smaller than the whole genome. 
When all species are represented in one plot in \citep{nasir17, caetano21}, 
including viruses, archaea, bacteria, eukaryotes, the scattering of the points (genomes)
in log-log scale does not follow one single straight line, but a curve.
Instead of treating the curve as consisting of multiple
piecewise linear regimes \citep{nasir17,caetano21}, another possibility is to use
a quadratic fitting function; admittedly, this does not provide an explanation for
the reason of deviation from the power-law trend, and it ignores some
fundamental differences in genome evolution in different domains.

One key lesson from the Heaps' law that the rate
of new knowledge gain does not linearly increase with the effort,
is very relevant to pan-genome studies \citep{tellelin} -- the number
of new genes discovered does not increase linearly with the number
of genome sequenced. A different kind of diminishing return, concerning 
the alignment of next-generation-sequencing reads when the read 
length is increased, is discussed in \citep{wli-read,stephens}. 

Besides the above work on protein fold superfamilies
in pan-genomes, Heaps' law, unlike Zipf's law \citep{parndt}, rarely 
appears in the literature concerning genomics. In \citep{gatherer},
when over-represented peptide substrings (over-represented k-mers
with k in a range of values)  are considered as words,
linear regression is used to fit the type-token scatter plots.  No attempt
was made to use the power-law function in \citep{gatherer}. In
\citep{som}, the fixed-length (k=3) codons are considered as words.
This definition of words would easily reach the vocabulary limit
when the text length is increased. Therefore, it is not surprising
that a different type-token relation is expected.

We have confirmed the result of some previous works that if tokens are randomly 
sampled from a dictionary (type pool), while type frequency (token per type)
follows the Zifp's law or some modification of the Zipf's law, the
resulting type-token plots, though may have a look of power-law/Heaps' law,
are actually curved \citep{bernhardsson,lv,font}. By simulation,
we also numerically determined the impact of finite dictionary size,
by removing the rare CRCPD types from consideration: the type-token 
plot becomes more curved with a reduction of dictionary size.
The previous attempts to derive the type-token relation, with finite 
or infinite dictionary size do not present the result in closed-form analytic
expression \citep{eliazar,lv,tunnicliffe}. Our quadratic regression 
provides a simple formula for an empirical description of the systematic
deviation from the Heaps' law.

The definition of ``type" or ``dictionary word", or in our case, Pfam domains,
is supposed to be given by the database, and synonyms are not merged.
For example, there are 13 protein domain names containing the prefix of zf-C2H2,
including the top ranking zf-C2H2 (5493 appearances in the genome) and
second ranking zf-C2H2\_6 (1609 appearance in the genome). The reason of
not combining potential synonyms is that it should be a feature of
a ``dictionary", and actually, Heaps' law partially reflects this lexical diversity.
If one would combine CRCPD that are ``similar" (up to certain degree),
the type-token plot and Heaps' law in human languages should also be modified
accordingly.

As argued in \citep{lynch}, a consequence of smaller population size
is that selection pressure is much less effective in removing individuals 
without a perfect fit for the given environment. As nonadaptive forces dominated,
genomes became less compact, with more redundancies, and inefficiencies were tolerated.
On the other hand, this near-neutral random walk in Wright's landscape allows
a species to reach a new adaptive peak, potentially higher than the previous one
\citep{wagner-book}. Microorganisms have populations several orders of magnitude larger 
than animals. The higher level of genome redundancy and complexity in animals,
versus the lower level of genome complexity in microorganisms,
is then expected. For a genome without redundancy, the number of token is
roughly equal to the number of types, and Heaps exponent is one. With redundancy,
exponent is expected to be less than  1.

It is suggested in \citep{caetano17,caetano21} that viral, archaeal and bacterial 
proteomics have a higher level of vocabulary growth than eukarya, as indicated
by the higher value of $\alpha$ exponents (around 0.8). The $\alpha$ exponent 
for eukarya genomes in \citep{caetano21} is only around 0.1, in 
contrast with our exponent $\alpha \sim 0.7$ within one genome. 
It is tempting to use the slope in linear regression to
rank different animal genomes in their level of duplication of protein domains.
However, the limited x-range makes the estimation of slope less reliable.
Also, when x-range is expanded, the linear (i.e., power-law when x and y are
not in log-scale) trend becomes quadratic, and there are three parameters
to characterize the type-token function, not just one. The similarity
of all three fitting parameters for the animal genomes in Fig.\ref{fig7-other-random}
make them more similar than different.

As a quantitative relationship, a Heaps' law like formula
can provide a quick order of magnitude estimation of human genome parameters 
\citep{li-hgp}. For example, if there is one copy of CRCPD per gene,
20000 genes in the human genome, and the Heaps' law exponent is 0.7, then 
20000$^{0.7} \approx$ 1000 protein domain types is estimated. 
With the factor C in Eq.\ref{eq-heaps}, we expect the
number to be further multiplied. For example, if C=5, we expect 5000 
protein domains (the actual number of Pfam domain types is 6839). 
Even though these numbers are  not exact, they 
provide a ballpark estimate. 

Despite the deviation from a perfect power-law for type-token 
plot, or in an alternative wording, that the power-law Heaps' law is only
technically correct in a limited range, there are other important lessons
in this study from the linguistic perspective. Just like human languages
with hundreds of new English words being created annually, 
albeit some of them are synonyms to the existing words,
new protein domains have been generated during evolution. 
Even if the new protein domains may not be completely new,
being at various degrees similar to the old ones, these may
not be that different from the new synonymous words in human languages.
Some of the seemingly non-novel new protein domains may continue to
evolve and eventually lead to a new meaning and a new structure.

\section*{Acknowledgment}
We would like to thank Patrick Villanueva for participating to the initial stage
of the project, Tom MacCarthy and Roman Samulyak for useful suggestions.

\newpage

\end{document}